# Qualitative Evaluation of LLM-Designed GUI

Bartosz Sawicki[1], Tomasz Leś[1], Dariusz Parzych[2],
Aleksandra Wycisk-Ficek[2], Paweł Trębacz[3], Paweł Zawadzki[1]

[1] Warsaw University of Technology, Faculty of Electrical Engineering
[2] Warsaw University of Technology, Strategic Analysis Department
[3] Warsaw University of Technology, Faculty of Architecture
bartosz.sawicki@pw.edu.pl

**Abstract.** As generative artificial intelligence advances, Large Language Models (LLMs) are being explored for automated graphical user interface (GUI) design. This study investigates the usability and adaptability of LLM-generated interfaces by analysing their ability to meet diverse user needs. The experiments included utilization of three state-of-the-art models from January 2025 (OpenAI GPT o3-mini-high, DeepSeek R1, and Anthropic Claude 3.5 Sonnet) generating mockups for three interface types: a chat system, a technical team panel, and a manager dashboard. Expert evaluations revealed that while LLMs are effective at creating structured layouts, they face challenges in meeting accessibility standards and providing interactive functionality. Further testing showed that LLMs could partially tailor interfaces for different user personas but lacked deeper contextual understanding. The results suggest that while LLMs are promising tools for early-stage UI prototyping, human intervention remains critical to ensure usability, accessibility, and user satisfaction.

## 1. Introduction

Graphical User Interfaces (GUIs) are fundamental to human-computer interaction. Traditionally, their design has relied on human expertise, but recent advancements in AI-driven design suggest that large language models (LLMs) can assist in automating this process. This study explores the qualitative aspects of LLM-generated GUIs, focusing on their usability and aesthetic.

Recent research highlights the potential of LLMs in generating graphical layouts from textual descriptions. For example, the TextLap framework tailors LLMs to assist in creating layouts based on user instructions. By leveraging a curated instruction-based dataset, TextLap effectively translates textual input into visual designs [16]. Similarly, the Low-code LLM framework introduces a novel interaction paradigm where users employ low-code visual programming to guide LLMs in complex tasks. This approach enhances the controllability and stability of generated outputs by allowing users to incorporate their preferences through a graphical interface [17].

To design GUIs that accommodate diverse user needs, integrating user modeling techniques is crucial. User modeling involves representing user characteristics—such as preferences, knowledge, and goals—to enable adaptive system behavior. Adaptive hypermedia systems, for instance, personalize content and navigation based on user models, improving usability and personalization.

Assessing the quality of graphical interfaces remains a challenge, as it involves subjective human evaluation. One initiative addressing this issue is WebDev Arena, developed by the LMArena team. This platform enables users to compare AI language models in web development by facilitating interactive testing and evaluation. Created by researchers from UC Berkeley SkyLab and LMSYS, LMArena supports crowdsourced assessments where users engage with AI chatbots, compare outputs, and provide ratings [18].

## 2. Graphical Interface for Humans

### 2.1. The role of GUI in building user experience

The user interface (UI) constitutes a key element of human-computer interaction, determining the way in which users communicate with digital systems. It serves as an intermediary layer between the human and the machine, with the primary goal of enabling efficient, intuitive, and ergonomic operation of information systems [1]. Modern UIs encompass a wide range of solutions—from traditional graphical user interfaces (GUIs) and touch interfaces to voice-controlled systems and augmented reality (AR) applications.

User interfaces have a significant impact on shaping the user experience (UX) by influencing levels of satisfaction, efficiency, and task performance [2]. The quality of a UI is closely linked to usability principles such as error minimization, clarity of communication, and alignment with the user's mental model [3]. In light of the rapid technological advancements, including those based on artificial intelligence, UIs also perform a mediating function: they simplify interactions and allow systems to be tailored to the specific needs of diverse user groups.

The evolution of user interfaces is driven both by technological progress and by changing user expectations. Designing an effective UI requires an interdisciplinary approach that combines insights from cognitive psychology, software engineering, and visual design. Future developments in this field are expected to emphasize intelligent interactions, adaptive interfaces, and the automation of interactive processes—opening new avenues for research into their impact on users and the overall effectiveness of digital systems.

The interface is to be a connection of the system with the user, therefore has to address not only the needs and expectations but also physical and perceptual limitations of different users and context of using the system. Its visual design and functionalities should be specifically tailored to a defined user group and conditions under which the device will be used (mobile / desktop device, indoor / outdoor) and even habits (in a hurry, while commuting). The interface is intended to facilitate action in various situations, most commonly resolving problems or addressing inconveniences. The key to build a user-friendly system is to clearly define problem categories and provide resolutions that encompass wide groups of user needs.

### 2.2. Methods of evaluating GUI

There are various methods for evaluating digital products. User-involved techniques are based on observing real system-human interactions using approaches derived from ethnography, sociology, and psychology, supported by contextual knowledge of market analysis. Research-based design has to process complex information on users' main characteristics, needs, and expectations.

One of the most influential approach to stay focused on the user while designing the system is a concept of persona, introduced by Alan Cooper in the 1990s. Personas are representing a construct of an idealised user, abstracted from information gathered through research, thus provide insights that ensure both high usability and user needs alignment [4].

Another popular approach in emphatic, human-centered design is a customer journey mapping, which enables designers to follow users' standard practices and learn how to minimise the negative and maximise the positive experience with the system. Early integration of UX research prevents designs from creating a system that ignores users' needs and fails to acknowledge users behavior patterns.

Usability is shaped by perception, which should not be constrained. The set of ten usability heuristics for user interface evaluation, developed by Jakob Nielsen and Rolf Molich in the 1990s, is a well-established method for identifying usability issues and is widely used in Human-Computer Interaction (HCI) research and UX/UI design. These heuristics cover key aspects of user-system interaction, such as system status visibility, consistency and standards, user control, and error minimisation [5]. The heuristic evaluation method enables the efficient detection of usability issues without requiring large groups of users, making it one of the most commonly used expert evaluation methods for interfaces [6].

Addressing the needs is essential for usability. The analysis of human needs is widely studied in psychological sciences. Numerous frameworks have been developed, these include one of the most well-known classifications—Abraham Maslow's five-level hierarchy of needs [7]—as well as the ancient division into seven groups of Hindu chakras [8]. Attempts to categorise human needs have resulted in detailed approaches [9] [10] [11] [12]. In 2000 another typology emerged, assuming that the degree of fulfilment is determined in the following areas: Autonomy, Beauty, Comfort, Community, Competence, Fitness, Impact, Morality, Purpose, Recognition, Relatedness, Security, Stimulation (with each of the 13 groups including subgroups, resulting in a total of 52 evaluation criteria) [13]. While this method is effective for assessing physical objects, it proves too extensive for interfaces, as many areas remain inactive.

Another method, more suited to this task, is the application of the Vitruvian triad: Commodity, Firmness, and Delight. In its modernised form, as presented by Capton [14], this approach defines specific categories and their corresponding optimal attributes. The primary categories include: Impartiality (Objectivity) of Form, Efficiency (Economy) of Function, Integrity (Propriety) of Meaning. The group of secondary categories consists of: Responsibility of Construction Design, Regard (Sympathy) for Context (Community), and Motivation (Conviction) of Spirit (Will). Translated into digital product, the Vitruvian categories, focus not only on the interface qualities such as pattern (e.g. overall readability of the layout, overall colour scheme), but also on the user: functionality and needs (e.g. large window, ease of using function keys), meaning and association (e.g. ability to review dialogue, easy access to request history), and attitudes (most essential / most easily accessible functionality).

The following research was based on a multi-faceted expert assessment with qualitative criteria derived from the above-mentioned studies in psychology approach and design fields. The score of the evaluation is the sum of the points attributed to rankings under these two methods: Nielsen's heuristics and architectural categories.

## 3. LLM Generates GUI

Within the conducted study, we aimed to evaluate the ability of LLM models to generate user interfaces by adopting three levels of complexity that reflect the increasing functional requirements of the system.

- The simplest level, designated as FixLine, involved creating a chat interface that enabled communication with a support system while offering the option to add photos and share locations.
- The second level, FixTeam, was designed for a technical team and provided functionalities to view reports, change their status, and add comments, thereby necessitating the implementation of more advanced interaction mechanisms.
- The most complex level, BoardPanel, was intended for managerial staff and provided access to overall statistics and detailed reports, which required data integration to support decision-making processes.

All generated interfaces were delivered as a single, self-contained HTML file that contained all necessary code (HTML, CSS, and JavaScript) and supported the use of any external libraries. These mock-ups are interactive, featuring basic functionality such as button interactions, dynamic content updates, and language switching. However, they do not include full backend integration, as their primary purpose is to evaluate how LLMs generate structured and partially functional UI designs based on textual descriptions.

We used Zero-shot learning prompting technique. It means a model understands and responds to tasks it has never seen before, purely by reasoning from broad knowledge.

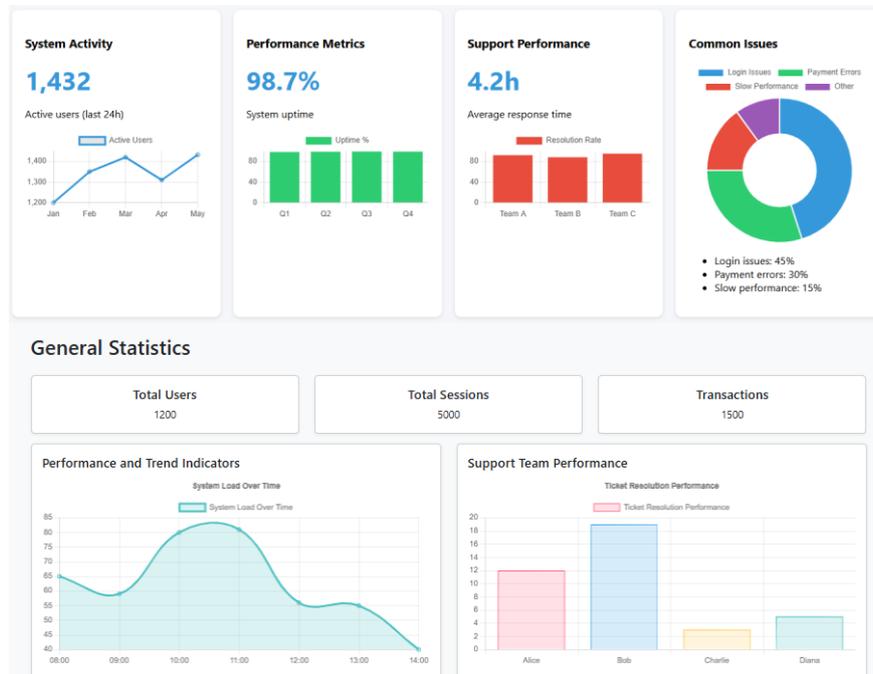

```
You are a programmer, an expert in graphical user interfaces for web
   applications. Create an interactive user interface mockup for the
 "BoardPanel" application main dashboard for high management persons.
    The application dashboard contains general statistics of system
activity; performance and trend indicators; support team performance;
most common problems reported. The application supports two language
versions: Polish and English. The mockup should be fully functional and
include sample data to allow full interface testing. Generate a single
HTML file containing all functionality (including JavaScript and CSS).
                    You may use external libraries.
```

Fig.1. Two different examples of interface mockup generated for management dashboard "BoardPanel"; Prompt used to produce dashboard mockup.

The process of generating interfaces with LLM begins with a precisely defined prompt that establishes the model's role as an expert in UI design and provides a complete set of functional requirements. This prompt clearly specifies what type of interface is to be generated while deliberately avoiding excessive details regarding aesthetics or code structure, thus allowing the model to determine the final appearance in accordance with current UI standards. Each task (FixLine, FixTeam, and BoardPanel) was associated with a distinct prompt tailored to its functional scope. However, within each task, all models received the same prompt to ensure comparability of results.

The experiments were conducted on three different models (Antropic Claude 3.5 Sonnet, OpenAI GPT o3-mini-high, and DeepSeek R1), generating three versions for each of the three interface types. In total 27 different mock-ups were generated for later evaluation. To ensure access to the latest available versions of each model, all experiments were performed on their respective native platforms. This approach allowed us to evaluate the models under real-world conditions, reflecting their intended deployment and usage scenarios.

**Experiment 1: OpenAI model o3-mini-high, date 2025-02-03**

Observations:
- Assigning the model a specialized role, such as a UI expert, enhances the quality of the generated results by ensuring alignment with best practices in interface design.
- The model tends to adopt a single solution path when given flexible instructions. For instance, if external libraries are mentioned as an option, the model will almost certainly incorporate them.
- Instructions placed at the end of the prompt carry higher priority. A structured approach, starting with a general problem introduction and ending with a clear expected outcome, yields better results.
- Large problems should be broken into smaller subproblems, with each prompt containing about 3–5 requirements. This balance ensures sufficient detail while avoiding shallow responses.
- As the only one of the three analysed models, o3-mini-high operated in reasoning mode, which gave us a clear advantage. Instead of requesting a complete solution upfront, guiding the model through a sequential problem-solving process improves code quality and readability.

**Experiment 2: DeepSeek model R1, date 2025-02-04**

Observations:

- While the model successfully produced interfaces based on a single prompt incorporating all specified conditions, deeper analysis revealed several functional issues:
  - Mocked data structures were generated but not correctly handled in the interface.
  - Essential interactive elements, such as message-sending buttons or list filters, did not trigger actions.
  - Language translation functionality was inconsistent, failing to translate all texts or modifying unintended elements.
- The experiments, conducted on February 3-4, 2025, coincided with high global traffic on DeepSeek, leading to frequent failures due to server overload. Successful requests experienced long response times compared to other models like GPT. Additionally, in one of thirty attempts, the output structure was so flawed that the code was entirely unusable.

- Despite some positive aspects, including DeepSeek's approach of first interpreting the prompt before generating output, inconsistencies in quality were observed. Each generated interface followed different conventions and coding practices, demonstrating variability in adherence to best programming standards.

**Experiment 3: Anthropic model Sonnet 3.5, date 2025-01-31**

Observations:
- The manufacturer does not provide a clear recommendation on which model will produce higher quality results. The Opus 3.0 model is positioned as more powerful, but the Sonnet 3.5 model was recently updated.
- Anthropic's platform offers a convenient runtime environment for web applications; however, it has limitations regarding external libraries. For this reason, some attempts resulted in critical errors.
- The model always produced syntactically correct code, which is a necessary condition for further testing.

## 4. Expert Evaluation

The anonymized interfaces were submitted for expert evaluation, preventing assessors from determining their origin. The expert evaluation followed the competent judges method [15] within a structured user experience assessment, using predefined criteria and rating scales in the following areas:
- Interface delivery: assessed as delivered/not delivered (0/1) in accordance with the prompt. Number of points reflects prompt complexity: chat prompt included 14 elements that could or could not be delivered, similarly: dashboard prompt included 8 elements, and technical prompt included 20 elements.
- Usability: evaluated using 10 Nielsen's Heuristics [5], on a three-point scale (meets criteria, does not meet criteria, difficult to determine).
- Accessibility: assessed as delivered/not delivered (0/1) based on text readability and contrast (sans-serif font, font size, clear contrast).
- Aesthetics & user needs: evaluated using 6 criteria derived from architectural categories on a five-point scale [14].
- Errors in generated content: classified as minor (hinders use), significant (substantially hinders use), or critical (prevents use of the system).

Each interface was evaluated using the above criteria and scales by independent experts (n=3) with background in psychology, sociology, cultural science, IT, user experience, architecture, and user perception research. The evaluation process included independent scoring by experts, followed by discussion of the assessment results. Though there were minor disagreements among the evaluators, there was a consensus in the scores for top performing models. After compiling the agreed evaluation results for the anonymized interfaces, the technical team informed the experts which model had generated the highest-rated materials.

The experts were unanimously in agreement on the evaluation of the chat interfaces and were mostly in agreement on the evaluation of the dashboard and technical interfaces. Interfaces with a similar rating (low or high) are shown in the table below (Tab.1).

According to the expert evaluation, the interfaces that best met the evaluation criteria are the following: chat generated by the DeepSeek R1 model; dashboard - by the Open AI GPT o3 model; technical - by Open AI GPT model o3).

The interfaces that met the evaluation criteria in the least are the following: chat generated by the Antrophic Claude 3.5 Sonnet model; dashboard - all generated by the Antrophic Claude 3.5 Sonnet model; technical - by DeepSeek R1 model and all generated by Antrophic Claude 3.5 Sonnet model.

All models showed high effectiveness in generating a well-structured spatial layout, e.g. proportional and aligned with format standards: a chat window in the form of dialogue box (some examples with text bubbles for each user), while a dashboard was composed of segmented sections with tables and charts. In some cases, the generated charts varied in type—most commonly line and bar charts, with pie charts appearing less frequently.

Tab.1 Final evaluation scores for each model (represented in percentage of maximum score possible for evaluated interface type). The highest ratings in a given category were marked in green.

| Model | chat (FixLine) | dashboard (BoardPanel) | technical (FixTeam) |
|---|---|---|---|
| Antrophic Claude a | 74% | 56% | 52% |
| Antrophic Claude b | 62% | 54% | 63% |
| Antrophic Claude c | 39% | 56% | 49% |
| DeepSeek a | 81% | 75% | 50% |
| DeepSeek b | 76% | 71% | 68% |
| DeepSeek c | 55% | 74% | 67% |
| GPT - OpenAI a | 71% | 75% | 77% |
| GPT - OpenAI b | 63% | 64% | 77% |
| GPT - OpenAI c | 51% | 78% | 86% |

Most interfaces used simple grey graphical elements with clearly marked CTAs (e.g., in one or two colours) and distinctly recognisable headers that guided the user through the system's narrative.

In most cases, some form of interactivity was observed, i.e. the interface accepted input text, assigned a registration number, switched the language from Polish to English, and allowed users to change the displayed data series on charts.

The most common errors varied depending on the type of element being evaluated. The chat interface included both critical mistakes (e.g. response triggered only by a specific word, incomplete Polish translation, missing CTA), and minor mistakes (e.g. lack of proportional layout, incorrectly coloured buttons, no option to edit or delete requests). Errors were noticed in dashboard interface as well (e.g. limited or non-functional table / chart formatting, no option to export reports, no help section or user guidance, lack of menu and log-in panel).

However, it was not clear to the experts how interactive each of the interface is meant to be, and consequently, how to assess elements that are acceptable in graphic mock-up but would be a critical mistake in a prototype, such as: non-functional buttons, lack of request registration; no error messages or field validation, no interaction with data (filtering, sorting).

The evaluation process demonstrated that the LLM-generated GUI have a limited usability for user experience research as with unclear status of the system (a mock-up, a wireframe, a prototype) it is impossible to clearly determine whether prompted conditions were met and the usability serves the user in accordance with the needs.

## 5. Adjusting interface to user needs

Despite its limits LLM-generated GUIs proved to be effective enough in building mock-ups in accordance with predefined criteria, thus we decided to continue the research with more complex requests in prompts.

In the second phase, we addressed the important research question—whether modern LLMs can adapt an interface to meet the needs of different user groups—by augmenting the base prompt. Case U1 corresponds to an interface tailored for a typical student interested in new technologies who uses a smartphone. Case U2 corresponds to an interface customized for a senior academic staff member who uses a smartphone.

The FixLine chat interface, generated by the OpenAI GPT-3-mini-high model, demonstrated notable adaptability in aligning design elements with the distinct preferences and requirements of these groups.

For U1 (students), the LLM produced an interface emphasizing dynamic aesthetics and efficiency: vibrant colors, compact navigation (all components in a single toolbar), smaller text input fields, and abbreviations (e.g., "EN/PL" for language selection). These choices reflect the tech-savvy, mobile-oriented habits of younger users, prioritizing quick interactions and visual appeal. However, potential trade-offs were observed, such as reduced readability of smaller fonts and ambiguity in abbreviated labels, suggesting a need for balanced minimalism.

In contrast, U2 (senior academics) featured a subdued, functional design: neutral colors, larger text for readability, explicit labels (e.g., "English/Polish"), and a spatially distributed layout. These adjustments cater to preferences for clarity, accessibility, and professional tone, critical for users less accustomed to dense digital interfaces. While functionally sound, the design lacked subtle enhancements (e.g., adjustable font sizes) to accommodate diverse accessibility needs.

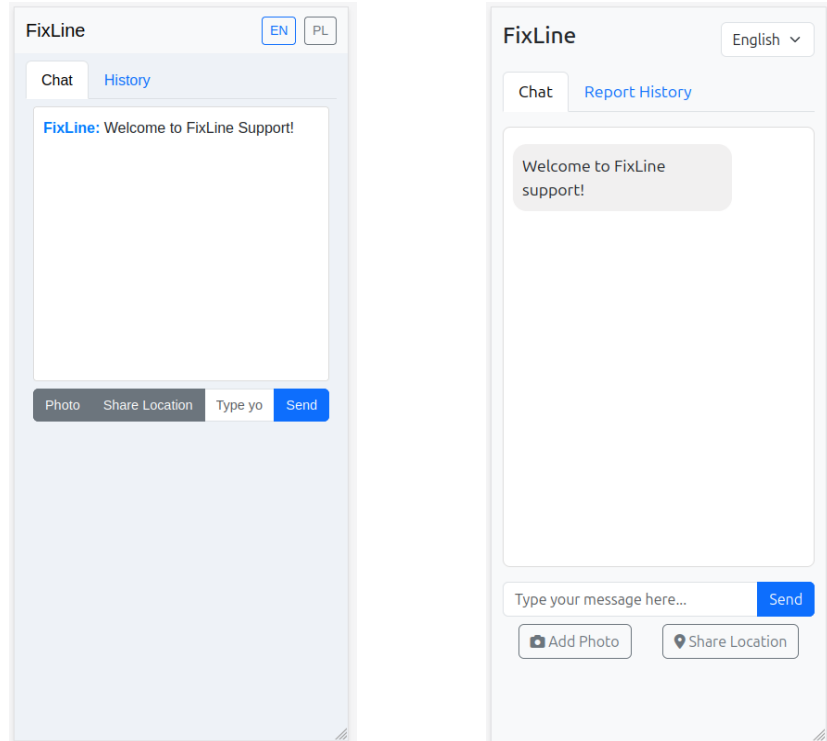

U1. "Adjust the interface for the convenience of a typical student who is interested in new technologies and uses a smartphone."

U2. "Adjust the interface for the convenience of a typical user who is a senior academic staff member and uses a smartphone."

Fig. 2. FixLine chat interface generated for two different types of users. U1 – technology student, U2 – senior academic.

The results underscore LLMs' capability to generate context-aware interfaces by interpreting user personas. However, limitations persist, particularly in addressing nuanced accessibility needs (e.g., customizable font sizes) and advanced functionalities. For instance, the absence of emoticons—a staple in youth-oriented apps—underscored a disconnect from habitual communication styles, revealing gaps in cultural contextualization. Future work should prioritize iterative usability testing with real users to validate design assumptions and refine prompts for granular personalization. While this phase confirms LLMs' potential as rapid prototyping tools, human oversight remains essential to bridge the gap between algorithmic outputs and human-centric design, ensuring interfaces align not only with functional requirements but also with users' cultural and behavioral nuances.

## 6. Conclusions

The evaluation demonstrated that the performance of the different LLMs varied according to the type of interface. All models presented impressive capabilities in creating interactive mock-ups, however "reasoning" models from OpenAI and DeepSeek slightly outperform Anthropic. Generating an accurate, user-friendly GUI requires improvement in two areas. First, the challenge of personalization is developing methods to accurately model and incorporate diverse user preferences into GUI generation. Secondly, the challenge concerning usability, i.e. ensuring that automatically generated interfaces are intuitive and meet the accessibility needs of various user groups. Addressing these challenges requires interdisciplinary research that integrates knowledge from artificial intelligence, human-computer interaction, and user experience design.

The evaluation highlighted challenges in using GenAI-generated interfaces for UX testing, particularly due to the ambiguous status of LLM-generated GUIs as either mockup or prototypes, leading to unclear user expectations. Our hypothesis assumed that all generated GUIs would adhere to the prompt as interactive mockups with sample data for interface testing; however, the produced mockups exhibited varying levels of interactivity—from minimal to nearly prototype-like.

Elements acceptable in mockups may be critical errors in prototypes, raising questions about interactivity criteria. This necessitates two possible UX testing strategies: treating the GUI as a graphic mockup useful for early design stages or as a prototype for later testing, provided it aligns precisely with design expectations. This discrepancy underscores the need for clearer interactivity requirements in the prompt. Additionally, refining assessment methodologies for LLM-generated GUIs would support their integration into design practices. While language models show potential in generating user-tailored interfaces, further advancements in user modelling, adaptive systems, and multimodal interactions are required for effective implementation.

## Acknowledgment

This research was supported by Warsaw University of Technology grant 'MyPW: Intelligent Campus Improvement System of the Warsaw University of Technology', No. CPR-IDUB/2/Z05/2025, implemented as part of STRATEG PW II call funded under the 'Excellence Initiative - Research University' program at the Warsaw University of Technology.